\documentclass[onecolumn,aps]{revtex4}
\usepackage{graphicx,psfrag,dsfont}
\usepackage{slashed,amssymb,amsmath}

\def\no{\noindent}
\def\bc{\begin{center}}
\def\ec{\end{center}}

\def\beq{\begin{equation}}
\def\eeq{\end{equation}}

\def\br{{\bf r}}
\def\bq{{\bf q}}

\def\bp{{\bf p}}

\newcommand{\nn}{\nonumber}

\begin{document}
\title{
Quantum Hall effect induced by electron-phonon interaction
}

\author{Andreas Sinner and Klaus Ziegler}
\affiliation{Institut f\"ur Physik, Universit\"at Augsburg, D-86135 Augsburg, Germany}

\begin{abstract}
\no
When phonons couple to fermions in 2D semimetals, the interaction may turn the system into an insulator. 
There are several insulating phases in which the time reversal and the sublattice symmetries are spontaneously broken. 
Examples are many-body states commensurate to Haldane's staggered flux model or to
lattice models with periodically modulated strain. We find that the effective field theories of these 
phases exhibit characteristic Chern-Simons terms, whose coefficients are related to the topological invariants of the
microscopic model. This implies that the corresponding quantized Hall conductivities characterize these insulating
states.
\end{abstract}

\maketitle

\section{Introduction}

The ground state of a many-body Hamiltonian may pass through a transition between phases 
with very different properties under a change of model parameters. This process is usually 
accompanied by the symmetry change of ground states on both sides of the
transition, the phenomenon known as the spontaneous symmetry breaking. The phase with the lower symmetry is characterized by the 
emergent macroscopic ordering of elementary building blocks of the system, technically captured by an order parameter. Phenomenological 
models describing the physics of the symmetry broken phase in terms of order parameters have proven very successful in the
description of various thermal and quantum phase transitions. In these models, critical parameters at which phase transitions occur
correspond to emergent non-trivial potential minima. They can be approached by variational techniques, which 
lead to a set of mean-field equations for all order parameters. Sometimes, the ground states of a system is not uniquely
determined by the order parameter alone, as they may have additional topological ordering~\cite{Wen1989,FradkinBook,Thouless1982}.
An important observation is that different phases of a degenerated ground state may be described by different topological field theories. 
Usually in condensed matter physics the emergence of the topological Chern-Simons excitations 
has not been considered in the broader context of symmetry breaking and phase transitions. To some extend, this is the case
for superfluid Helium, where $^3{\rm He}-{\rm A}$ phase exhibits also a spontaneous time reversal symmetry breaking among other spontaneously broken 
symmetries~\cite{Volovik1988a,Volovik1988b,VollhardtBook}.

\vspace{1mm}
\no 
The phase of the wave function can play a crucial role for the properties of the related physical system \cite{aharonov-bohm59,berry84}. 
In particular, if this phase is associated with a topological invariant through the winding number, it may
describe robust macroscopic properties. Examples in condensed matter physics are the quantized Hall conductivity 
of electrons in 2d~\cite{FradkinBook,Thouless1982,Haldane1988,Halperin1982,Froehlich1991,Yakovenko1990,Zee1995} 
and in quasi 1d structures~\cite{Trauzettel2012}. Another example is 
the quantized transverse conductance of the supercurrent in the
$^3{\rm He}-{\rm A}$-phase of superfluid liquid Helium~\cite{Volovik1988a,Volovik1988b,VollhardtBook},
and quantized conductivities of different anomalous Hall effects~\cite{Nagaosa2010,Hankiewicz2005,Buttiker2012}.
In general, topological structures in 2+1-dimensional gauge field theories are 
related to Chern-Simons terms, where the coefficient in front of the latter is related 
to a topological invariant~\cite{Redlich1984,Jackiw1984,Fradkin1994,Dunne1999,Hughes2008,Vishwanath2012}.

\vspace{1mm}
\no 
The idea that dynamical lattice degrees of freedom can be treated as effective gauge fields is not new~\cite{Zohar2016,Heidari2019}. 
Here we study a system of monochromatic phonons minimally coupled to Dirac fermions in 2D. 
As the electron-phonon coupling strength increases the system should approach
a transition into the structurally different phase. Our intention is to identify possible gapped phases, the corresponding 
quasiparticles and observables. The connection with tight-binding models will help us to link these phases to different
lattice distortions, which lead to the formation of spectral gaps with broken time reversal and sublattice symmetries.
To emphasize the similarity of phonons and gauge fields we consider the $E^{}_1$-optical phonon mode of a honeycomb 
lattice which represents individual vibrations of the sublattices in opposite directions~\cite{Basko2008,Amorim2016}. The number of 
modes is therefore naturally restricted to 2 and there is no time-like component of the effective gauge field.
This resembles the popular in electrodynamics {\textquotedblleft light cone\textquotedblright} gauge fixing $\hat n\cdot\vec A=0$, with $\hat n$ 
denoting a unit vector pointing into the time direction~\cite{Wen1989}. It was shown in~\cite{Phonon16,Phonon19} that in such 
systems phonon Chern-Simons excitations can be generated.

\section{The model}
\label{sec:MicModel}

The properties of tight-binding Hamiltonians  which describe the motion of electrons on the honeycomb lattice are well known and 
investigated in details in the existing literature~\cite{Wallace1947,Semenoff1984}. In sublattice representation it reads
\begin{equation}
\label{eq:tbh00}
H^{}_{\rm el} = \sum_{\br,\br'}h^{}_{1;\br\br'}c^\dagger_\br\cdot\sigma^{}_1c^{}_{\br'} + \sum_{\br,\br'}h^{}_{2;\br\br'}c^\dagger_\br\cdot\sigma^{}_2c^{}_{\br'} = 
\sum_{\br,\br'} c^\dagger_\br\cdot H^{}_{0;\br\br'} c^{}_{\br'} ,
\end{equation}
with the Pauli matrices $\sigma^{}_\mu$ reflecting the sublattice degree of freedom. The hopping matrix elements are
\begin{eqnarray}
h^{}_{1;\br\br'}  &=& -\frac{t}{2}\sum_{i=1,2,3} \left(\delta^{}_{\br',\br+{\bf a}^{}_i} + \delta^{}_{\br,\br'+{\bf a}^{}_i} \right),\\
h^{}_{2;\br\br'}  &=& -\frac{t}{2i}\sum_{i=1,2,3}\left(\delta^{}_{\br',\br+{\bf a}^{}_i} - \delta^{}_{\br,\br'+{\bf a}^{}_i} \right),
\end{eqnarray}
where $t$ denotes the hopping amplitude and $\{{\bf a}^{}_i\}^{}_{i=1,2,3}$ are the nearest neighbor basis vectors on a lattice. 
The Fourier components of the hopping matrices read $h^{}_{1}(\bp) = -t\sum_{i=1\cdots3}  \cos(\bp\cdot {\bf a}^{}_i)$ and $h^{}_{2}(\bp) =-t\sum_{i=1\cdots3}  \sin(\bp\cdot {\bf a}^{}_i)$, and 
which gives for the kernel of the Hamiltonian 
\begin{equation}
\label{eq:kern}
H^{}_0(\bp)=h^{}_1(\bp)\sigma^{}_1+h^{}_2(\bp)\sigma^{}_2.
\end{equation}
The tight-binding Hamiltonian Eq.~(\ref{eq:tbh00}) is symmetric under time reversal $H^T_{\rm el} = H^{}_{\rm el}$, which translates into the Fourier space as
\begin{equation}
\label{eq:TRS}
H^\ast_{\rm 0}(-\bp) = H^{}_{\rm 0}(\bp),
\end{equation}
with the complex conjugation acting only on the sublattice space. 
A spectral gap appears in the Hamiltonian in the form of the Dirac mass parameter~\cite{Semenoff1984}
$\displaystyle m\sum_\br c^\dag_\br\sigma^{}_3c^{}_\br$, and  does not affect the property~(\ref{eq:TRS}). 

\vspace{1mm}
The dynamics of two-component (in-plane) monochromatic phonons is governed by  
\begin{equation}
\label{eq:PhHam}
H^{}_{\rm ph} = \sum_{\mu=1,2}\sum_{\br}\frac{1}{2}(P_{\mu;\br}^2 + \omega^2 A_{\mu;\br}^2), 
\end{equation}
with dispersionless frequency  $\omega$ and operators of canonical momentum ($P$) and position ($A$)
\begin{eqnarray}
P^{}_{\mu;\br} &=& i\sqrt{\frac{\omega}{2}} \sum^{}_{\bq} \left[b^{}_{\mu,\bq}e^{i\bq\cdot\br} - b^\dag_{\mu,\bq}e^{-i\bq\cdot\br}\right],\\
A^{}_{\mu;\br} &=& \frac{1}{\sqrt{2\omega}} \sum^{}_{\bq} \left[b^{}_{\mu,\bq}e^{i\bq\cdot\br} + b^\dag_{\mu,\bq}e^{-i\bq\cdot\br} \right],
\end{eqnarray}
expressed in terms of bosonic operators $b^{}_{\mu,\br}(b^\dag_{\mu,\br})$ which annihilate (create) a phonon at position $\br$. 
Phonons interact with electrons via attachment to the lattice bonds
\begin{equation}
\label{interaction00}
H^{}_{\rm I} = U\sum_{\mu=1,2}\sum_{\br}c^\dagger_\br\cdot A_{\mu;\br}\sigma^{}_\mu c^{}_{\br},
\end{equation}
whereas the coupling parameter $U$ can in principle be time dependent and describe a periodic driving. 
In doing this one can open up new directions toward physics of emergent dynamical Floquet topological phases~\cite{lindner11,cayssol13}
or non-equilibrium superconductivity~\cite{Babadi2016,Babadi2017,Demler19}. 

\subsection{Functional integral representation}

The coherent state functional integral representation of the zero temperature
partition function corresponding to the model Hamiltonian 
\begin{equation}
H = H^{}_{\rm el} + H^{}_{\rm ph} + H^{}_{\rm I}
\end{equation}
reads~
\begin{equation}
{\cal Z}  = \int{\cal D}[\psi^\dag,\psi,A]\exp\{-{\cal S}[\psi^\dag,\psi,A]\},
\end{equation}
with the Euclidean lattice action
\begin{equation}
{\cal S}[\psi^\dag,\psi,A] = {\cal S}^{}_{\rm el}[\psi^\dag,\psi] + {\cal S}^{}_{\rm ph}[A] + {\cal S}^{}_{\rm I}[\psi^\dag,\psi,A].
\end{equation}
The elements of the action are
\begin{eqnarray}
{\cal S}^{}_{\rm el}[\psi^\dag,\psi] &=& \sum_{\br,\br'} \int d\tau~ \psi^\dag_{\br}(\tau)[\partial^{}_\tau\sigma^{}_0 + H^{}_{0;\br\br'}]\psi^{}_{\br'}(\tau), \\
\label{eq:PhAct}
{\cal S}^{}_{\rm ph}[A] &=& \frac{1}{2}\sum_{\mu=1,2}\sum_{\br} \int d\tau~ \left\{[\partial^{}_\tau A^{}_{\mu;\br}(\tau)]^2 + [\nabla^{}_\br A^{}_{\mu;\br}(\tau)]^2 + \omega^2A^2_{\mu;\br}(\tau) \right\}, \\
{\cal S}^{}_{\rm I}[\psi^\dag,\psi,A] &=& U\sum_{\mu=1,2}\sum_{\br}\int d\tau~ A^{}_{\mu;\br}(\tau)\psi^\dag_{\br}(\tau) \sigma^{}_\mu \psi^{}_{\br}(\tau).
\end{eqnarray}
Here $\psi=(\psi^{}_1,\psi^{}_2)$ represents a two-component time and position dependent Grassmann spinor with the component index referring to the sublattice.  
In this representation, the model resembles the standard two-dimensional Su-Schrieffer-Heeger (SSH) model of electron-phonon interaction~\cite{SSH1979} with phonons coupling 
to the electron currents~\cite{Zoli2004,Zoli2005}. In contrast to the conventional SSH model our phonons are dispersionless, which allows us to make the crucial approximation by 
dropping the kinetic term $(\partial^{}_\tau A^{}_\mu)^2+(\nabla^{}_\br A^{}_\mu)^2$.
The physical picture of this approximation consists in neglecting the slow dynamics of heavy lattice ions due to thermal fluctuations close to the 
absolute zero of temperature in comparison to the induced dynamics due to the interaction with band electrons. This approximation is justified only if 
phonons have a relatively large spectral gap, which is the case for optical phonon branches of honeycomb lattices~\cite{Amorim2016,Ferrari2007,Sanders2013,Jorio2012}.

\subsection{Effective low-energy model}

The low-energy part of the spectrum of $H^{}_{0}$ defined in Eq.~(\ref{eq:kern}) is characterized by two Dirac cones at ${\bf K}^{}_\pm$, 
which represent Fermi quasiparticles with different chiralities and introduces an additional discrete degree of freedom into the model. 
This requires operating in the space of $4\times4$ complex matrices $\Sigma^{}_{ij} = \sigma^{}_{i}\otimes\sigma^{}_{j},\; i,j=0,1,2,3$. 
In total there are 15 traceless matrices corresponding to a particular representation of the group of $SU(4)$-transformations and a 4-dimensional unit matrix $\Sigma^{}_{00}$. 
For any complex $4\times4$-matrix ${\cal Q}$ there is a unique decomposition ${\cal Q}={\cal Q}^{ij}\Sigma^{}_{ij}$. 
Using this notation the effective low-energy  Hamiltonian reads
\begin{equation}
\label{eq:Weyl}
H^{}_{\rm W}(\bp) = p^{}_1\Sigma^{}_{01} + p^{}_2\Sigma^{}_{32}.
\end{equation}
A reasonable translation of the property~(\ref{eq:TRS}) for the low-energy Hamiltonian is  
$\Sigma^{}_{02}H^{\rm T}_{\rm W}(-\bp)\Sigma^{}_{02}=H^{}_{\rm W}(\bp)$ for the time reversal and $\Sigma^{}_{33}H^{}_{\rm W}(\bp)\Sigma^{}_{33}=-H^{}_{\rm W}(\bp)$ 
for the sublattice (or chiral) symmetry~\cite{Ludwig1994}.

\vspace{1mm}
While in the lattice model defined in the previous paragraph, the phonon field is attached to the lattice bonds, 
in low-energy approximation it is attached to each sublattice and accounts for the scattering between both Dirac nodes. 
This is dictated by the properties of the $C^{}_{6v}$ group~\cite{Basko2008,Phonon19} 
and models the $E^{}_1$-in-plane optical modes. The spectrum of these modes reveals a weak alteration over the 
entire Brillouin zone~\cite{Ferrari2007,Sanders2013,Amorim2016}, which makes it possible to model them in 
the form of dispersionless monochromatic lattice vibrations. 
The effective low-energy Hamiltonian which describes the Dirac fermions coupled to the in-plane phonons in 2D reads
\begin{equation}
\label{eq:Ham}
{\cal H} = i\vec\partial\cdot\vec j + \vec A\cdot\vec \Pi, 
\end{equation}
where we assembled two particular subsets with two $\Sigma$-matrices each to vectors $\vec j = \{\Sigma^{}_{01},\Sigma^{}_{32}\}$ and 
$\vec\Pi=\{\Sigma^{}_{13},\Sigma^{}_{20}\}$. The set  $\vec j$ couples to the kinetic energy of fermions, while $\vec\Pi$ to the two-component phonon field $\vec A$. 
The corresponding Euclidean action becomes:
\begin{equation}
\label{eq:Action}
{\cal S}[A,\psi^\dag,\psi] = \frac{1}{2g}\vec A\cdot\vec A + \psi^\dag\cdot\left[\partial^{}_\tau\Sigma^{}_{00} + {\cal H} \right]\psi,
\end{equation}
where $\psi=(\psi_{11},\psi^{}_{12},\psi^{}_{21},\psi^{}_{22})^{\rm T}$ are the complex Grassmann fields 
with the first index referring to the sublattice and the second to the respective Dirac point, and $\cal H$ is given in Eq.~(\ref{eq:Ham}). 
The coupling constant $g\sim  U^2/\omega^2$ is related to the inverse frequency of monochromatic phonons~\cite{Phonon16,Basko2008}. 
The correlation functions of currents can be obtained from the generating functional 
\begin{equation}
\label{eq:GenFunc}
{\cal W}[{\cal A}]=\log\langle\exp\{-{\cal S}[{\cal A}]\}\rangle^{}_{\psi^\dag\psi,A}
\end{equation}
by repeating variations with respect to the fields ${\cal A}=\{\vec\alpha,\vec\beta\}$,
${\cal S}[{\cal A}]$ is the action~(\ref{eq:Action}) augmented by an auxiliary source term 
\begin{equation}
\psi^\dag[\vec\alpha\cdot\vec j + \vec\beta\cdot\vec\Pi]\psi = \vec\alpha\cdot\vec J^{\rm intra}+\vec\beta\cdot\vec J^{\rm inter}
\end{equation}
with introduced intranodal and internodal current operators 
\begin{equation}
 \vec J^{\rm intra}=\vec j=\{\Sigma^{}_{01},\Sigma^{}_{32}\},\;\,\; \vec J^{\rm inter}=\vec\Pi=\{\Sigma^{}_{13},\Sigma^{}_{20}\}.
\end{equation}
The averaging operator $\langle\cdots\rangle^{}_{\psi^\dag\psi,A}$ denotes the functional integration over all degrees of freedom. 
Field $\alpha$ is the external gauge field while field $\beta$ could be an external mechanical modulated strain field. 
Particularly interesting for symmetry broken phases are non-vanishing average currents ($\mu=1,2$)
\begin{eqnarray}
\label{eq:ElCur}
\bar J^{\rm intra}_{\mu,r} &=& -\left.\frac{\delta}{\delta\alpha^{}_{\mu,r}}\right|_{{\cal A}=0} {\cal W}[{\cal A}], \\
\label{eq:ThCur}
\bar J^{\rm inter}_{\mu,r} &=& -\left.\frac{\delta}{\delta \beta^{}_{\mu,r}}\right|_{{\cal A}=0} {\cal W}[{\cal A}]. 
\end{eqnarray}

\vspace{1mm}
\no
Following the procedure developed in Ref.~\cite{Phonon19}, we integrate out the bosonic field 
$\vec A$, which creates a four-fermion interaction term. The latter can be decoupled anew by 
4$\times$4 matrix fields $\cal Q$:
\begin{equation}
\frac{1}{2g} A^2_\mu + A^{}_\mu\psi^\dag\Pi^{}_\mu\psi\rightarrow 
-\frac{g}{2} \left(\psi^\dag\Pi^{}_\mu\psi\right)^2 = \frac{g}{2}{\rm tr}
\left(\Pi^{}_\mu\psi\psi^\dag\right)^2 \rightarrow \frac{1}{2g}{\rm tr}{\cal Q}^2_\mu 
+i\psi^\dag{\cal Q}^{}_\mu\Pi^{}_\mu\psi,
\end{equation}
where the summation over $\mu=1,2$ is understood and the decoupling field decomposes as 
${\cal Q}^{}_\mu={\cal Q}^{ij}_\mu\Sigma^{}_{ij}$. The full action in this representation reads
\begin{equation}
{\cal S}[{\cal Q},\psi^\dag,\psi] = \frac{1}{2g}{\rm tr}{\cal Q}^2_\mu 
+\psi^\dag\cdot\left[G^{-1}_0 + i{\cal Q}^{}_\mu\Pi^{}_\mu\right]\psi,\;\;\; 
G^{-1}_0 = \partial^{}_\tau\Sigma^{}_{00} + i\vec \partial\cdot\vec j
 .
\end{equation}
Now we can integrate out the fermions to arrive at the bosonic action
\begin{equation}
\label{eq:BosAct}
{\cal S}[{\cal Q}] = \frac{1}{2g}{\rm tr}{\cal Q}^2_\mu - {\rm tr}\log\left[G^{-1}_0 
+ i{\cal Q}^{}_\mu\Pi^{}_\mu \right].
\end{equation}

\section{Saddle-point analysis}

\subsection{Effective potential}

In simplest mean field approximation we replace quantum fields ${\cal Q}^{}_\mu$ in action~(\ref{eq:BosAct}) by corresponding 
spatially uniform classical background fields $M^{}_\mu$ 
\begin{equation}
\label{eq:EffPot}
V^{}_{\rm eff} = \frac{1}{2g}{\rm tr}M^2_\mu - \log\det\left[G^{-1}_0 +  iM^{}_\mu\Pi^{}_\mu \right].
\end{equation}
To provide a spectral gap, the gap parameter must couple to a matrix which does not commute with the low-energy Hamiltonian $H^{}_{\rm W}$ in Eq.~(\ref{eq:Weyl}). 
Technically this prevents singularities on the real axis in the Green's functions. This restricts our freedom to 
$M^{}_1 = (\Delta\Sigma^{}_{01} + m\Sigma^{}_{20})/2$ and $M^{}_{2} = (\Delta\Sigma^{}_{32} - m\Sigma^{}_{13})/2$, such that 
\begin{equation}
\label{eq:OrdPar}
iM^{}_\mu\Pi^{}_\mu = m\Sigma^{}_{33}  + \Delta\Sigma^{}_{12},
\end{equation}
which indeed anticommutes with $H^{}_{\rm W}$. There are two further matrices which anticommute with $H^{}_{\rm W}$: $\Sigma^{}_{03}$ and $\Sigma^{}_{22}$. 
It is easy to see though, that the self-consistent equations for the respective order-parameters are unstable. 
For $\Sigma^{}_{03}$ this is demonstrated in Ref.~\cite{Phonon19}, while the case of $\Sigma^{}_{22}$ is ruled out in the 
Appendix~\ref{app:Delta2}. 

\begin{figure*}[t]
\includegraphics[width=7cm]{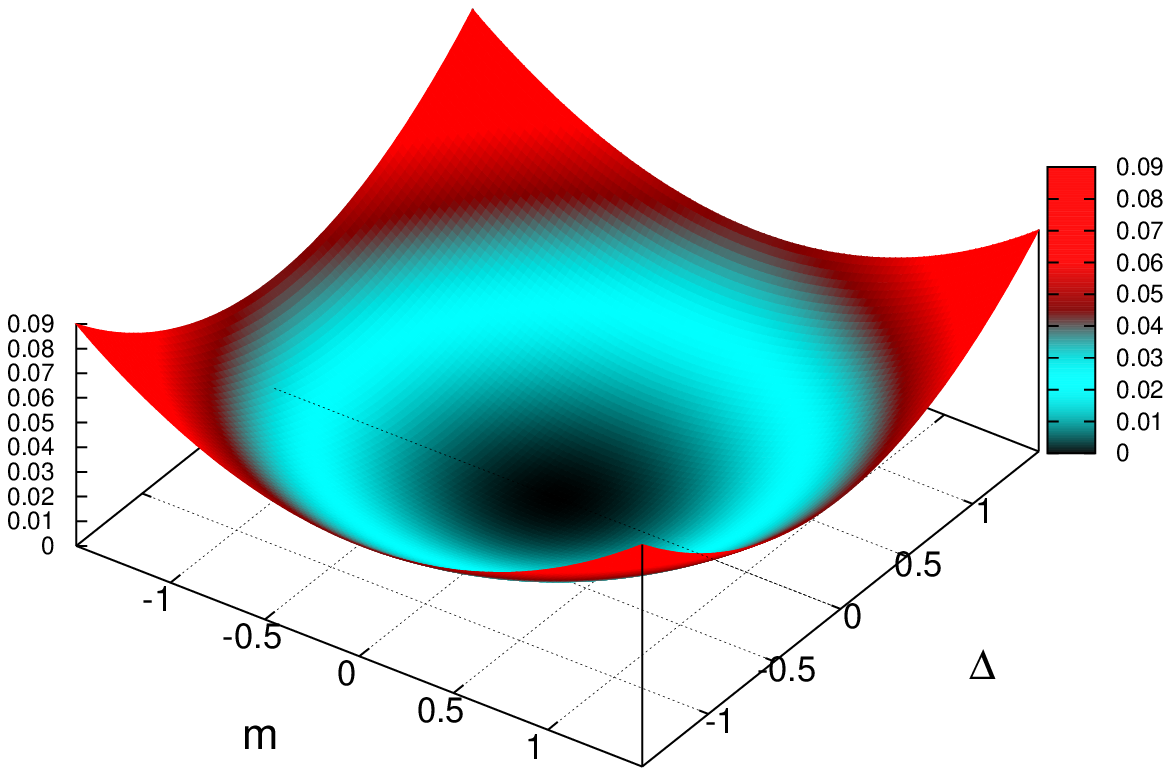} 
\includegraphics[width=7cm]{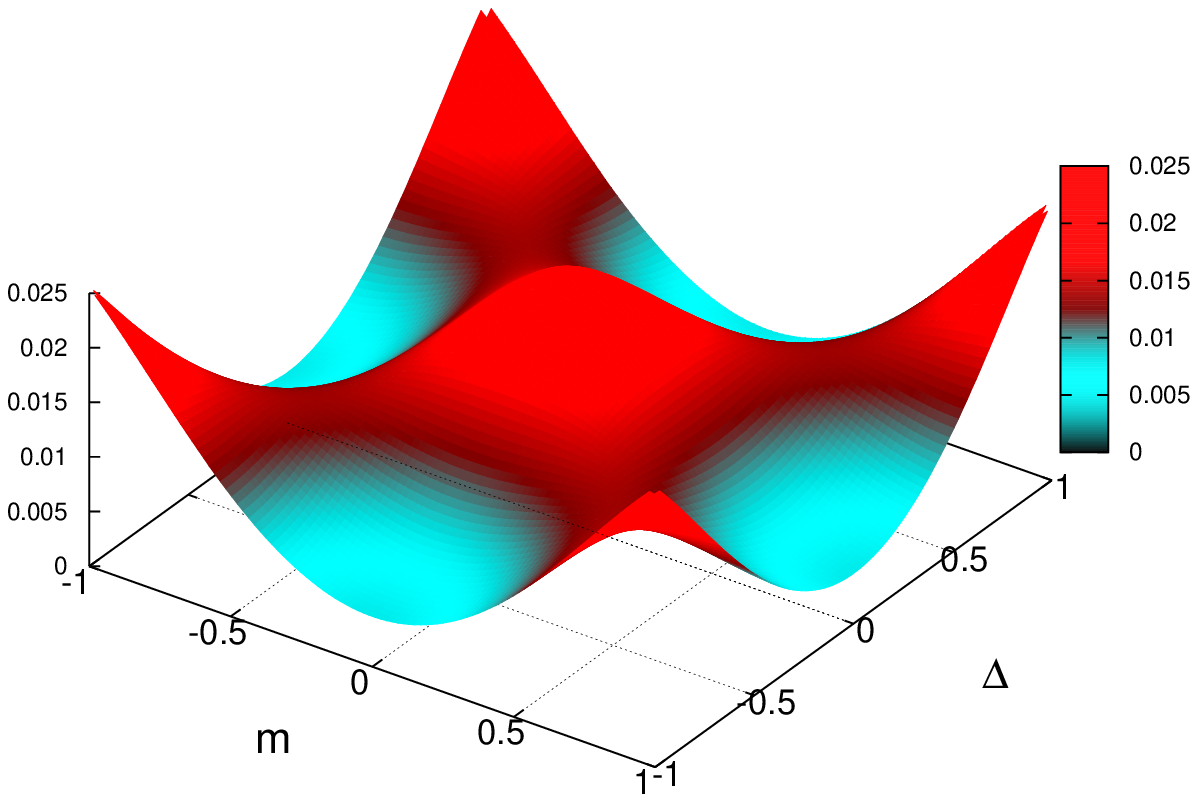} 
\caption{
Left panel: Low-energy landscape of the mean-field free energy for $g/g^{}_c<1$ with the only minimum at $m=0=\Delta$.
Right panel: Low-energy landscape of the mean-field free energy with visible barriers between stable phases. 
The model parameter are chosen to give $\gamma=g/g^{}_c\sim 3.5$. 
}
\label{fig:Pot}
\end{figure*}

\vspace{1mm}
The minus sign between both terms in Eq.~(\ref{eq:EffPot}) suggests a competition between them. For small $g$ the first term dominates and the effective 
potential has the shape of a convex hull as it is shown in the left panel of Figure~\ref{fig:Pot}. As the interaction becomes larger, both terms become
comparable in size until the second term destroys the convexity of the potential. The determinant can be readily evaluated giving 
\begin{equation}
\log\det\left[G^{-1}_0 +  iM^{}_\mu\Pi^{}_\mu \right] = 
\int\frac{d^3Q}{(2\pi)^3}~\log\left[\Delta^4 + 2\Delta^2 (Q^2-m^2) + (Q^2+m^2)^2 \right], 
\end{equation}
where $Q^2=q^2_0+\bq^2$. The argument of the logarithm is not indifferent to the interchange of $m$ and $\Delta$, which suggests the anisotropy of the potential. 
All integrals diverge and we need to perform the integrations up to a spherical cutoff. The potential landscape plotted in the right panel of Figure~\ref{fig:Pot} 
reveals a high, yet discrete symmetry. Visually one observes four angular minima, corresponding to two stable phases with degenerated vacua 
separated from each other by potential walls. On the ridge of the potential one recognizes a fourfold degenerated saddle point like structure, 
which corresponds to the unstable states with strong tendency to decay in one of both stable phases.
The $\Delta$-- and the $m$--phase are associated with extra terms in the tight-binding Hamiltonian~(\ref{eq:tbh00})
in real space representation:
\begin{equation}
H^{}_{\rm TB;{\bf r}{\bf r}'} + \Delta\sigma_2 \cos({\bf G}\cdot{\bf r})\delta_{{\bf r},{\bf r}'},
\label{haldane_ham}
\end{equation}
where ${\bf G}={\bf K}_+-{\bf K}_-$, and
\begin{equation}
H^{}_{\rm TB;{\bf r}{\bf r}'}+im\sigma_3\sum_{j=1}^3(\delta_{{\bf r}',{\bf r}+{\bf c}_j}
-\delta_{{\bf r}',{\bf r}-{\bf c}_j}),
\end{equation}
$\{{\bf c}_j\}_{j=1,2,3}$ denoting nearest neighbor basis vectors on one sublattice \cite{Haldane1988,Hill2011}.
The $\Delta$--phase is characterized by a strain field which modulates with the nodal
wave vector ${\bf G}$ (modulated strain phase), while the $m$--phase is a flux phase that is commensurate with one 
sublattice (Haldane phase). Both phases are subject to time reversal and sublattice symmetry breaking.

\begin{figure*}[t]
\includegraphics[width=7cm]{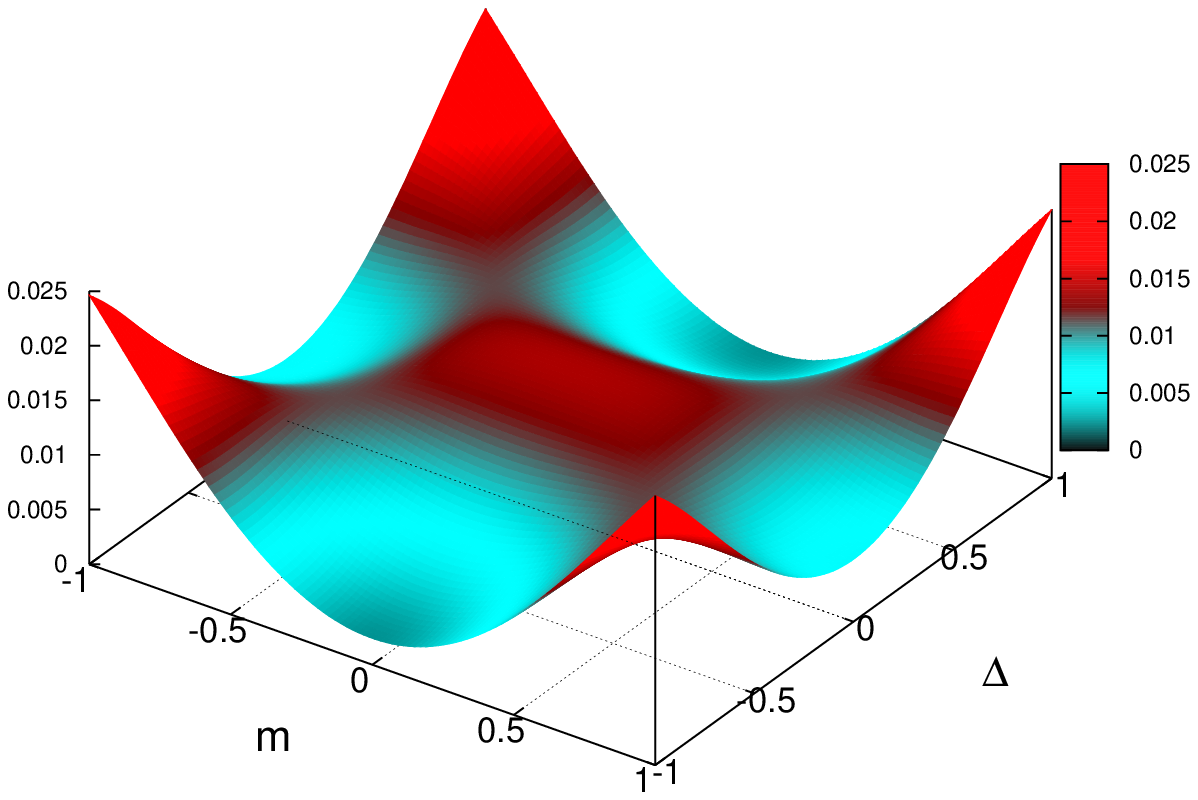} 
\includegraphics[width=7cm]{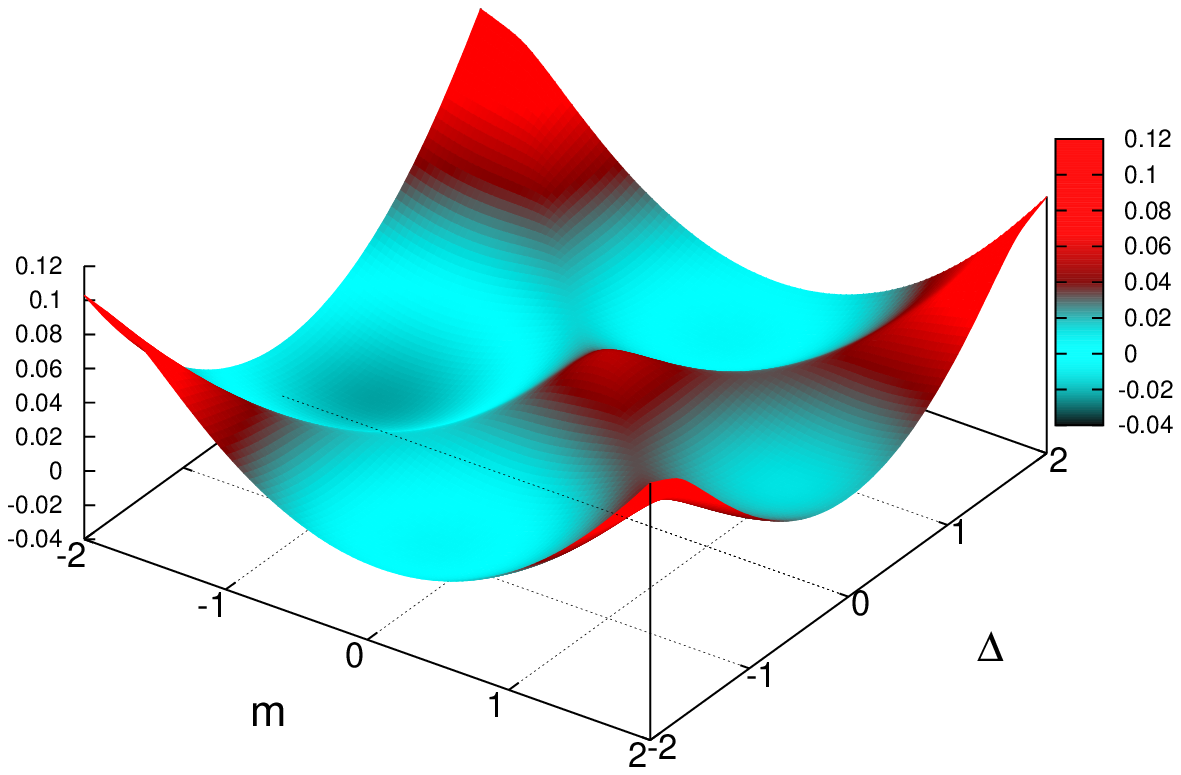} 
\caption{
Two examples for broken symmetry between both ground states. 
Upper row, left panel: Two perturbations which anticommute with the bare electronic Hamiltonian 
make the $\Delta$-phase less pronounced compared to the $m$-phase. They couple 
to matrices $\Sigma^{}_{03}$ (staggered chemical potential) and $\Sigma^{}_{22}$. 
Upper row, right panel: The situation which favors the $\Delta$-phase can be realized with 
two further perturbations which anticommute with the bare electronic Hamiltonian.
The couple to $\Sigma^{}_{33}$ and $\Sigma^{}_{12}$ phases. Depending on the sign 
of the perturbation, the degeneracy between the two symmetric minima is lifted as well.
}
\label{fig:SymBrPot}
\end{figure*}

\vspace{1mm}
The minima corresponding to both phases are equally deep, i.e. the phases are evenly likely to emerge. A natural question to ask is therefore, 
if this observation points to a possible symmetry between both phases. Then there would be a unitary transformation which maps one vacuum on another and vice versa. 
It is indeed possible to construct a unitary transformation which transforms $\Sigma^{}_{12}$ on $\Sigma^{}_{33}$, i.e. 
\begin{equation}
T^\dag\Sigma^{}_{12}T = \Sigma^{}_{33}.
\end{equation}
Explicitly we get 
\begin{equation}
T^\dag = \frac{1}{2} 
\left( 
\begin{array}{cccc}
1 & -i & 1 & -i \\
1 &  i & 1 &  i \\
1 & -i &-1 &  i \\
1 &  i &-1 & -i
\end{array}
\right)
, \;\; 
T = \frac{1}{2} 
\left( 
\begin{array}{cccc}
1 & 1 & 1 & 1 \\
i &-i & i & -i \\
1 & 1 & -1 & -1 \\
i & -i & -i & i
\end{array}
\right).
\end{equation}
The transformation $T^\dag$ is not a symmetry of the bare electronic Hamiltonian,
nor of the mean-field Hamiltonian, but because of its unitarity leaves the eigenvalues unchanged. 

\vspace{1mm}
Technically, the simplest way to break the symmetry between both stable ground states 
is by a perturbation which anticommutes with the bare electronic Hamiltonian Eq.~(\ref{eq:Weyl}). This property guaranties, that the determinant 
of the perturbed Hamiltonian remains rotationally invariant which greatly simplifies the calculations. If the perturbation mimics the order parameter, i.e. it couples either to 
$\Sigma^{}_{12}$ or $\Sigma^{}_{33}$ matrix, playing the role of the remanent density of respective fields, then the strain phase (order parameter $\Delta$) is energetically 
more preferable and even the $Z^{}_2$-symmetry between the both minima is violated, as it can be seen in the right panel of Figure~\ref{fig:SymBrPot}. The two other possible 
perturbations couple to either $\Sigma^{}_{03}$ or $\Sigma^{}_{22}$ and favor the flux phase (order parameter $m$). In this case the $Z^{}_2$-degeneracy of the ground state 
remains present, cf. left panel of Figure~\ref{fig:SymBrPot}. 
On the lattice the corresponding contributions to the Hamiltonian are: if $\Sigma^{}_{03}$ is chosen 
\begin{eqnarray}
\delta H^{(1)}_{\rm p} = u^{}_1\sigma^{}_3\delta^{}_{\br,\br'},
\end{eqnarray}
and if $\Sigma^{}_{22}$
\begin{eqnarray}
\delta H^{(2)}_{\rm p} = i u^{}_2\sigma_2 \sin({\bf G}\cdot{\bf r})\delta_{{\bf r},{\bf r}'}, 
\end{eqnarray}
where ${\bf G}={\bf K}_+-{\bf K}_-$.

\subsection{Solution of the saddle-point equations}

A more detailed picture of the symmetry breaking follows from the solutions of mean field equations. 
They are obtained if potential~(\ref{eq:EffPot}) is varied with respect to $M^{}_\alpha$: 
\begin{equation}
\label{eq:SPE}
M^{}_\alpha = ig\Pi^{}_\alpha \left[G^{-1}_0 + iM^{}_\mu\Pi^{}_\mu \right]^{-1}_{rr}.
\end{equation}
Inserting the order parameter matrix in (\ref{eq:SPE}), performing the frequency integration from $-\infty$ to $+\infty$ and the radial integral up to the 
cutoff $\Lambda$, related to the band width we get 
\begin{equation}
m\pm\Delta = \gamma (m\pm \Delta) \left[\sqrt{1+(m\pm\Delta)^2} - |m\pm\Delta|\right],
\end{equation}
where $\gamma = g\Lambda/2\pi$ and $m$ and $\Delta$ are rescaled in units of $\Lambda$. First we notice that $m=\Delta$ might be a solution 
for both $+$, as well as trivially for $-$ sign. This also includes $m=0=\Delta$ case. For $+$ sign, the non-trivial solutions 
are extracted from 
\begin{equation}
1 = \gamma [\sqrt{1+4m^2}-2|m|], 
\end{equation}
which exists only if the interaction strength exceeds a critical value of the order of inverse band width $\gamma = g/g^{}_c\geqslant1$, $g^{}_c=2\pi/\Lambda$
\begin{equation}
\label{eq:Coex1}
|m| = |\Delta| = \frac{\gamma^2-1}{4\gamma}\Theta(\gamma-1).
\end{equation}
This solution corresponds to the state which is visible on the ridge of the phase separating walls in Figure~\ref{fig:Pot}. 
This state is stable only classically (mean-field) and is destroyed by quantum fluctuations as it will be shown below. 
Further solutions are there for $m\neq0$, $\Delta=0$ and for $m=0$, $\Delta\neq0$. For $\Delta$-phase we get 
\begin{equation}
\label{eq:Coex}
|\Delta| = \frac{\gamma^2-1}{2\gamma}\Theta(\gamma-1),
\end{equation}
and the same for $m$-phase. These solutions correspond to the four angular minima visible in Figure~\ref{fig:Pot}. They turn out to be stable against quantum fluctuations as well.

\section{Quantum fluctuations} 

Expanding Eq.~(\ref{eq:BosAct}) to the second order in the fluctuations around ground states of respective phases we get
\begin{equation}
\label{eq:Gauss}
{\cal S}^{}_G[{\cal Q}] = {\rm tr}\left\{ \frac{1}{2g}{\cal Q}^2_\mu - 
\frac{1}{2}G{\cal Q}^{}_\alpha\Pi^{}_\alpha G{\cal Q}^{}_\beta\Pi^{}_\beta\right\},
\end{equation}
which can be brought further into the conventional quadratic form 
\begin{equation}
\label{eq:Gauss2}
{\cal S}^{}_G[{\cal Q}] = \int\frac{d^3P}{(2\pi)^3}~
\left( 
\begin{array}{c}
\vec{\cal Q}^{}_1\\
\vec{\cal Q}^{}_2
\end{array}
\right)^{\rm T}_P
\left( 
\begin{array}{cc}
 {\bf A} & {\bf B} \\
 {\bf C} & {\bf D}
\end{array}
\right)^{}_P
\left( 
\begin{array}{c}
\vec{\cal Q}^{}_1\\
\vec{\cal Q}^{}_2
\end{array}
\right)^{}_{-P}
\end{equation}
with the vector fields
\begin{equation}
\vec{\cal Q}^{}_{i} = \left({\cal Q}^{00}_i,{\cal Q}^{01}_i,{\cal Q}^{02}_i,{\cal Q}^{03}_i, 
{\cal Q}^{10}_i,{\cal Q}^{11}_i,{\cal Q}^{12}_i,{\cal Q}^{13}_i,{\cal Q}^{20}_i,{\cal Q}^{21}_i,
{\cal Q}^{22}_i,{\cal Q}^{23}_i,{\cal Q}^{30}_i,{\cal Q}^{31}_i,{\cal Q}^{32}_i,{\cal Q}^{33}_i\right), 
\;\; i=1,2.
\end{equation}
The kernel matrix of the quadratic form, (i.e. inverse bosonic propagator matrix) 
\begin{equation}
{\bf \Pi}^{-1}(P) = 
\left( 
\begin{array}{cc}
 {\bf A} & {\bf B} \\
 {\bf C} & {\bf D}
\end{array}
\right)^{}_P
\end{equation} 
is a 32$\times$32 matrix, i.e., each matrix block itself is a 16$\times$16 matrix. In accord with the intuitive picture drawn from the analysis of the effective potential, 
the momentum-independent terms in Gaussian action (\ref{eq:Gauss2}) (which are commonly called the Proca matrices) of both 
{\it m}-phase and  $\Delta$-phase are indeed strictly positive, while in the mixed phase between them with $m=\Delta\neq 0$ found in Eq.~(\ref{eq:Coex1}) 
there are negative eigenvalues, rendering this phase unstable, cf. Appendix~\ref{app:Proca}. Because the Proca matrix has no zero eigenvalues the kernel matrix in~(\ref{eq:Gauss2}) can be 
inverted even for zero momentum. An astonishing fact about the Proca matrices is that some of their elements sum up to multiples of $\gamma^{-1}$, cf. Appendix~\ref{app:Proca}. 
In calculating the Hall conductivities, Proca matrices guarantee the convergence of the functional integral but their elements do not appear in the ultimate result.

\vspace{1mm}
In exploring the structure of the gradient expansion we restrict our analysis to the linear terms only. 
In the usual three-dimensional electrodynamics it is referred to as the induced Chern-Simons terms~\cite{Redlich1984,Jackiw1984,Fradkin1994,Dunne1999}. 
These terms have several important properties: they are odd under time-reversal and because they are not associated with any length scale they represent pristine 
topological excitations with both scale and conformal invariance. Because of these qualities they are believed to represent the effective field theoretical description 
of the quantum Hall effect~\cite{Zee1995}. In our formalism things are getting more involved. At first glance the frequency dependent part of what is going to become the Chern-Simons tensor reads
\begin{equation}
{\cal S}^{(0)}_{CS} = \int\frac{d^3P}{(2\pi)^3}~p^{}_0 {\cal Q}^{ab}_{\mu,P}{\cal Q}^{\alpha\beta}_{\nu,-P}~{\cal P}^{\mu\nu}_{ab|\alpha\beta},
\end{equation}
where 
\begin{eqnarray}
{\cal P}^{\mu\nu}_{ab|\alpha\beta} &=& \frac{i}{2}{\rm Tr}\int\frac{d^3Q}{(2\pi)^3}~G(Q)\Sigma^{}_{ab}\Pi^{}_\mu G(Q)G(Q)\Sigma^{}_{\alpha\beta}\Pi^{}_\nu \\
&=& \frac{i}{32\pi}{\rm Tr}\left\{\hat\Sigma\left(\Sigma^{}_{ab}\Pi^{}_\mu\Sigma^{}_{\alpha\beta}\Pi^{}_\nu - \Sigma^{}_{\alpha\beta}\Pi^{}_\nu\Sigma^{}_{ab}\Pi^{}_\mu \right)\right\},
\end{eqnarray}
with $\hat\Sigma = {\rm sgn}(m)\Sigma^{}_{33}$ in $m$-phase and $\hat\Sigma = {\rm sgn}(\Delta)\Sigma^{}_{12}$ in $\Delta$-phase. 
For spatial components we get at first
\begin{equation}
{\cal S}^{}_{CS} = \int\frac{d^3Q}{(2\pi)^3}~ {\cal Q}^{ab}_{\mu,P}{\cal Q}^{\alpha\beta}_{\nu,-P}~ p^{}_i \cdot {\cal T}^{i,\mu\nu}_{ab|\alpha\beta},
\end{equation}
where 
\begin{eqnarray}
{\cal T}^{i,\mu\nu}_{ab|\alpha\beta} &=& \frac{1}{2}{\rm Tr}\int\frac{d^3Q}{(2\pi)^3}~G(Q)\Sigma^{}_{ab}\Pi^{}_\mu G(Q)j^{}_i G(Q)\Sigma^{}_{\alpha\beta}\Pi^{}_\nu \\
&=& - \frac{1}{32\pi}{\rm Tr}\left\{ 
\hat\Sigma\left( 
\Sigma^{}_{ab}\Pi^{}_\mu j^{}_i\Sigma^{}_{\alpha\beta}\Pi^{}_\nu - \Sigma^{}_{\alpha\beta}\Pi^{}_\nu j^{}_i \Sigma^{}_{ab}\Pi^{}_\mu
\right)
\right\}.
\end{eqnarray}
A decisive simplification of the above expressions becomes possible with the knowledge of certain distinct features of $\Sigma$-matrices. 
The full set of 16 $\Sigma$-matrices subdivides into four subsets of four matrices each, 
with three of them anticommuting among each other and the fourth commuting with all. 
Calling them $\varLambda^{a}_{i}$ with ${a=1,2,3,4}$ and ${i=0,1,2,3}$ and imposing the normalization condition ${\rm Tr}[\varLambda^a_0\varLambda^a_1\varLambda^a_2\varLambda^a_3]=4$ 
for all $a$, the subsets become 
\begin{eqnarray}
\displaystyle \varLambda^1 = \{i\Sigma^{}_{00},-\Sigma^{}_{01},-\Sigma^{}_{32},-\Sigma^{}_{33}\},  & & \displaystyle \varLambda^2=\{-i\Sigma^{}_{21},\Sigma^{}_{20},-\Sigma^{}_{13},\Sigma^{}_{12}\}, \\
\displaystyle \varLambda^3 = \{\Sigma^{}_{30},-i\Sigma^{}_{31},i\Sigma^{}_{02},-i\Sigma^{}_{03}\}, & & \displaystyle \varLambda^4 = \{-i\Sigma^{}_{11},\Sigma^{}_{10},\Sigma^{}_{23},-\Sigma^{}_{22}\}.
\end{eqnarray}
One recognizes that $\varLambda^1$ and $\varLambda^2$ include vectors $\vec j$ and $\vec\Pi$ of the model defined in~(\ref{eq:Ham}). 
Linear terms of gradient expansion of the action~(\ref{eq:Gauss}) in both phases can be brought into the form structurally similar to the conventional
Chern-Simons terms. The crucial difference is that rather than in terms of single components of $\cal Q$-fields, the Chern-Simons-like structures appear in terms of
\begin{equation}
\label{eq:Lfield}
{\bf \Lambda}^{a}_{i} = {\rm Tr}[\varLambda^a_{i}({\cal Q}^{}_{\mu}\Pi^{}_\mu)] 
\end{equation}
for each set $\varLambda^a$. In {\it m}-phase it resembles the conventional Chern-Simons terms, giving one Chern-Simons term for each set 
\begin{equation}
\label{eq:CSm}
{\cal S}^{m}_{CS} = S^{}_m\epsilon^{}_{ijk}{\bf \Lambda}^a_i\cdot i{\partial}^{}_j{\bf \Lambda}^a_k,
\end{equation}
where the dot-product implies the integration over the entire 3d space-time, $i,j,k=0,1,2$, and $S^{}_m={\rm sgn}(m)/64\pi$. 
The fact that in Eq.~(\ref{eq:CSm}) the derivatives with respect to spatial degrees of freedom appear alongside with the temporal 
can be interpreted as the restoration of the full gauge invariance, which formally lacks in action Eq.~(\ref{eq:Action})~\cite{Phonon16}. 

\vspace{1mm}
For the $\Delta$-phase we get two morphologically different terms. The first one resembles the generalized multi-field 
Chern-Simons term~\cite{Vishwanath2012}
\begin{equation}
\label{eq:CSDelta}
{\cal S}^{\Delta}_{CS} = S^{}_\Delta\epsilon^{}_{ijk}K^{}_{ab} {\bf \Lambda}^a_i\cdot i{\partial}^{}_j{\bf \Lambda}^b_k,
\end{equation}
where upper indices run only over $a,b=1,2$ and diagonal elements of symmetric tensor $K$ are zero. Effectively it connects the  
intranodal and internodal currents with each other. 
The two remaining matrix sets $\varLambda^{3}$ and $\varLambda^{4}$ interact with each other in an unexpected way 
\begin{equation}
\label{eq:CSDelta2}
\tilde{\cal S}^{\Delta}_{{CS}} = S^{}_\Delta K^{}_{ab} 
\left[{\bf \Lambda}^a_3\cdot i{\partial}^{}_i{\bf \Lambda}^b_i - {\bf \Lambda}^b_i\cdot i{\partial}^{}_i{\bf \Lambda}^a_3\right],
\end{equation}
where $i=0,1,2$ and $a,b=3,4$. The two terms (\ref{eq:CSDelta}) and (\ref{eq:CSDelta2}) are hardly discussed in context of induced Chern-Simons terms~\cite{Dunne1999,Zee1995}. 
From Eq.~(\ref{eq:CSm}) and (\ref{eq:CSDelta}) it is possible to construct a generalized Chern-Simons-like tensor for both phases. 
For this we retain only two Pauli sets $\Lambda^1$ and $\Lambda^2$ which are connected to the effective model~(\ref{eq:Ham}) in the  
sense that they contain both intra- and internode currents. With $\alpha,\beta=1,2$ we can combine them in one single expression
\begin{equation}
\label{eq:CS}
{\cal S}^{}_{CS} = \epsilon^{}_{ijk}T^{}_{cd} {\bf \Lambda}^c_i\cdot i{\partial}^{}_j{\bf \Lambda}^d_k,
\end{equation}
where the diagonal elements of the symmetric tensor $T^{}_{cc}={\rm sgn}(m)/64\pi$ and non-diagonal ones $T^{}_{cd}={\rm sgn}(\Delta)/64\pi$. 
Together with the Proca matrices of respective phases given in Appendix~\ref{app:Proca}, Eq.~(\ref{eq:CS}) is used to calculate the topological 
response function specific to each identified phase.

\subsection{Generalized Hall response}

We introduce a generalized current-current correlation function
\begin{equation}
\label{eq:CurCor}
\langle J^{c}_{x,\mu}J^{d}_{y,\nu} \rangle^{}_{\psi^\dag\psi,A} = \left.\frac{\delta}{\delta{\cal A}^d_{y,\nu}} \frac{\delta}{\delta{\cal A}^c_{x,\mu}}\right|_{{\cal A}=0}{\cal W}[{\cal A}], 
\end{equation}
where $\langle \cdots \rangle^{}_{\psi^\dag\psi,A}$ denotes the normalized functional integration
with respect to the action in Eq.~(\ref{eq:Action}) and ${\cal W}[{\cal A}]$ the generating functional of current-current correlator functions 
defined in Eq.~(\ref{eq:GenFunc}). The lower current indices refer to the 
position and time coordinates of the local current on the lattice $(x,y)$ and to their spatial 
direction $(\mu,\nu)$. Upper indices $(a,b)$ refer to the type of the current and distinguishes 
between inter- and intranode currents. The index combination $c=d=1$ corresponds to the correlator of usual intranodal currents, 
\begin{equation}
\label{eq:JJcor}
\langle (\psi^\dag j^{}_\mu\psi)^{}_x (\psi^\dag j^{}_\nu\psi)^{}_y \rangle^{}_{\psi^\dag\psi,A} \rightarrow  
\frac{1}{4\gamma^2}\langle {\rm Tr}\left( j^{}_\mu\Pi^{}_a{\cal Q}^{}_a\right)^{}_x {\rm Tr}\left(j^{}_\nu\Pi^{}_b{\cal Q}^{}_b\right)^{}_y \rangle^{}_{\cal Q}
\rightarrow  \frac{1}{4\gamma^2}\langle {\bf \Lambda}^1_{\mu,x} {\bf \Lambda}^1_{\nu,y}\rangle^{}_{\cal G},
\end{equation}
where in the second term we brought it into the $\cal Q$-matrix representation and introduced in the third $\bf\Lambda$-notation according to Eq.~(\ref{eq:Lfield}). 
The operator $\langle\cdots\rangle^{}_{\cal Q}$ denotes the average over the non-linear action Eq.~(\ref{eq:BosAct}) while  $\langle\cdots\rangle^{}_{\cal G}$ over
its Gaussian approximation Eq.~(\ref{eq:Gauss}). In the same manner we recognize that $c=d=2$-case represents 
the correlator internodal currents. The response in the $\Delta$-phase with $c=1$, $d=2$ and vice versa represents correlators over 
products of an intra- and an internodal current flowing in perpendicular directions. With all four terms we get for Eq.~(\ref{eq:CurCor})
\begin{equation}
\label{eq:CurCor2}
\langle J^{c}_{x,\mu}J^{d}_{y,\nu} \rangle^{}_{\psi^\dag\psi,A} \approx\frac{1}{4\gamma^2}\langle{\bf \Lambda}^c_{\mu,x} {\bf \Lambda}^d_{\nu,y}\rangle^{}_{\cal G}.
\end{equation}
The physical content of this correlator is simple: We create a local current density of the sort $c$ at position $x$ in direction $\mu$. Then 
we measure the local current of type $d$ at position $y$ in direction $\nu$. This suggests that not only correlators between pairs of intra-(inter-)node currents are possible, 
but also mixed correlators of the type $\langle J^{\rm intra}_{x,\mu}J^{\rm inter}_{y,\nu} \rangle$. 
With explicit knowledge of Proca matrices for each phase and of the generalized Chern-Simons term Eq.~(\ref{eq:CS}), the correlator of each two currents can be 
evaluated giving to linear order in gradient expansion, c.f. Appendix~\ref{app:JJ} 
\begin{equation}
\label{eq:npKubo}
\frac{1}{4\gamma^2}\langle {\bf \Lambda}^c_{i,r} {\bf \Lambda}^d_{j,r^\prime}\rangle^{}_{\cal G} = 32T^{}_{cd} i\epsilon^{}_{ijk}\partial^{}_k\delta(r-r^\prime). 
\end{equation}
In deriving this expression we used the fact that some elements of the Proca matrices sum up to multiples of $\gamma^{-1}$.
This means that while the presence of the Proca matrix is absolutely indispensable in order to carry out the functional integration over the $\cal Q$-fields, 
its elements do not appear in the correlator. 

\vspace{1mm}
The current-current correlator is related to the Hall conductivity via the Kubo formula~\cite{allen}
\begin{eqnarray}
\bar\sigma^{cd}_{\mu\nu} = \lim_{\omega\to0}\frac{2\pi}{\omega} \int d^3x \left[e^{-i\omega(x^{}_0-y^{}_0)} 
-1\right]\langle J^{c}_{x,\mu}J^{d}_{y,\nu} \rangle^{}_{\psi^\dag\psi,A}. 
\end{eqnarray}
Going through the algebra we ultimately find
\begin{equation}
\label{eq:GenKubo}
\bar\sigma^{cd}_{\mu\nu}= 64\pi T^{}_{cd}\epsilon^{}_{\mu\nu}.
\end{equation}
Diagonal elements of (\ref{eq:GenKubo}) give the quantized response ${\rm sgn}(m)\epsilon^{}_{\mu\nu}$ in the $m$-phase and the two off-diagonal components 
give the quantized response ${\rm sgn}(\Delta)\epsilon^{}_{\mu\nu}$ of the $\Delta$-phase. 
$c=d=1$ corresponds to the usual intranodal Hall conductivity of two Dirac cones, while $c=d=2$ represents the average of two internodal currents, 
which gives the internodal Hall conductivity. The response in the $\Delta$-phase with $c=1$, $d=2$ and vice versa represents the averages over 
products of an intra- and an internodal current flowing in perpendicular directions, which too has a quantized response.

\section{Discussions} 

The low-energy behavior of free fermions on the honeycomb lattice is determined by separated spectral nodes. Because of these
one can distinguish between currents directly linked to each node, i.e. the intranodal currents, and those
between the nodes, i.e. the internodal currents. The system of such fermions coupled to the in-plane phonons can be formally interpreted as 
a gauge field theory. At sufficiently large electron-phonon interaction strength a transition into a structurally different lattice phase takes place, 
as visualized in Figure~\ref{fig:Pot}.
We identify two distinct ground states with order parameters which break both the time reversal and the sublattice symmetries.
The effective phonon model exhibits the Chern-Simons-like terms in both phases with coefficients related to quantized Hall conductivities.
A similar effect of spontaneous time-reversal symmetry breaking was previously discussed for a 
time-periodic driven quantum system, where also a topological state with non-zero Chern numbers
was observed \cite{lindner11}. The similarity of the two cases is reflected by the mean-field Hamiltonian (\ref{haldane_ham}),
which agrees with the effective Floquet Hamiltonian in low energy approximation \cite{lindner11,cayssol13}.
This indicates that the time-dependent phonon field plays a similar role as a time-periodic driving field. 

\vspace{1mm}
\no
For the experimental observation of the proposed effects we suggest the following setup: 
the measurement could be performed on a suspended tightened graphene sample which must be sufficiently 
flat in order to rule out disorder due to ripples and out-of-plane phonon modes. The in-plane phonon modes could be 
generated mechanically by applying modulated strain with variable frequency on the sample and by adjusting carefully the 
strain frequency to the energy of the $E^{}_1$- mode. Estimations based on ab-initio calculations 
vary between $0.15-0.2$eV~\cite{Basko2008,Sanders2013}, which is well within the range of infrared Raman scattering technique~\cite{Jorio2012}.
In order to measure the Hall conductivities a weak electric gradient should be applied to the sample.
In the case of the $\Delta$-phase, both electric and strain fields should be applied in the same direction 
to create the proposed quantized response.

\section*{ACKNOWLEDGMENTS}
\no
Illuminating discussions with John Chalker, Bertrand Halperin, Eugene Demler, and Alexei Tsvelik are gratefully acknowledged.
We thank Aditi Mitra for drawing our attention to connections between time-periodically
driven quantum systems and systems with electron-phonon interaction.
This research was supported by the grants of the Julian Schwinger Foundation for Physics Research.

\appendix

\section{Exclusion of $\Delta^{}_2$ parameter}
\label{app:Delta2}

\no
Using the ansatz
\begin{equation}
 M^{}_1 = \frac{\Delta^{}_1}{2}\Sigma^{}_{01} -i \frac{\Delta^{}_2}{2}\Sigma^{}_{31} + \frac{m}{2} \Sigma^{}_{20}, \;\;\;   
 M^{}_2 = \frac{\Delta^{}_1}{2}\Sigma^{}_{32} -i \frac{\Delta^{}_2}{2}\Sigma^{}_{02} - \frac{m}{2} \Sigma^{}_{13}, 
\end{equation}
with which 
\begin{equation}
M = iM^{}_\mu\Pi^{}_\mu = m\Sigma^{}_{33}  + \Delta^{}_1\Sigma^{}_{12} + \Delta^{}_2 \Sigma^{}_{22}. 
\end{equation}
we get to the set of mean-field equations
\begin{equation}
\label{eq:SPE1}
M^{}_\alpha = ig\Pi^{}_\alpha\left[G^{-1}_0 + M \right]^{-1}_{rr}.
\end{equation}
Projecting both sides on the subspaces of the composite order parameter we get for each $\alpha$ 
\begin{eqnarray}
    m        &=& ~ \frac{g}{2} {\rm Tr}\Sigma^{}_{33}\left[G^{-1}_0 + M\right]^{-1}_{rr},\\
\Delta^{}_1  &=& ~ \frac{g}{2} {\rm Tr}\Sigma^{}_{12}\left[G^{-1}_0 + M\right]^{-1}_{rr},\\
\Delta^{}_2  &=& - \frac{g}{2} {\rm Tr}\Sigma^{}_{22}\left[G^{-1}_0 + M\right]^{-1}_{rr},
\end{eqnarray}
which becomes after inverting the inverse propagator matrix
\begin{eqnarray}
\label{eq:MF1}
     m       &=& ~     2mg      \int\frac{d^3Q}{(2\pi)^3} ~ \frac{q^2_0 + q^2 + m^2 - \Delta^2_1 - \Delta^2_2}{m^4 + 2m^2(q^2_0 + q^2 - \Delta^2_1 - \Delta^2_2) + (q^2_0+q^2+\Delta^2_1+\Delta^2_2)^2}, \\
\label{eq:MF2}     
\Delta^{}_1  &=& ~2\Delta^{}_1g \int\frac{d^3Q}{(2\pi)^3} ~ \frac{q^2_0 + q^2 - m^2 + \Delta^2_1 + \Delta^2_2}{m^4 + 2m^2(q^2_0 + q^2 - \Delta^2_1 - \Delta^2_2) + (q^2_0+q^2+\Delta^2_1+\Delta^2_2)^2}, \\
\label{eq:MF3}
\Delta^{}_2  &=& - 2\Delta^{}_2g \int\frac{d^3Q}{(2\pi)^3} ~ \frac{q^2_0 + q^2 - m^2 + \Delta^2_1 + \Delta^2_2}{m^4 + 2m^2(q^2_0 + q^2 - \Delta^2_1 - \Delta^2_2) + (q^2_0+q^2+\Delta^2_1+\Delta^2_2)^2}. 
\end{eqnarray}
If there are any mutually independent non-trivial solutions for $\Delta^{}_1$ and $\Delta^{}_2$ then they follow from 
\begin{eqnarray}
\label{eq:MF21}     
    1  &=& ~2g \int\frac{d^3Q}{(2\pi)^3} ~ \frac{q^2_0 + q^2 - m^2 + \Delta^2_1 + \Delta^2_2}{m^4 + 2m^2(q^2_0 + q^2 - \Delta^2_1 - \Delta^2_2) + (q^2_0+q^2+\Delta^2_1+\Delta^2_2)^2}, \\
\label{eq:MF31}
    1  &=& - 2g \int\frac{d^3Q}{(2\pi)^3} ~ \frac{q^2_0 + q^2 - m^2 + \Delta^2_1 + \Delta^2_2}{m^4 + 2m^2(q^2_0 + q^2 - \Delta^2_1 - \Delta^2_2) + (q^2_0+q^2+\Delta^2_1+\Delta^2_2)^2}, 
\end{eqnarray}
i.~e. only one of them can be true and they cannot be fulfilled together. The two possibilities are 
\begin{eqnarray}
\label{eq:poss1} 
\Delta^{}_1 = 0,\;\;\;  \Delta^{}_2\neq0; \\
\label{eq:poss2} 
\Delta^{}_2 = 0,\;\;\;  \Delta^{}_1\neq0.
\end{eqnarray}

\vspace{1mm}
The scenario of Eq.~(\ref{eq:poss2}) is discussed in the main text. Here we consider the scenario of Eq.~(\ref{eq:poss1}), i.~e. $\Delta^{}_{1}=0$ and $\Delta^{}_2 = \Delta$ is finite. 
The system of equations (\ref{eq:MF1})-(\ref{eq:MF3}) reduces to 
\begin{eqnarray}
    m   &=&   2   m   g \int\frac{d^3Q}{(2\pi)^3} ~\frac{q^2_0+q^2+m^2-\Delta^2}{m^4 + 2m^2(q^2_0+q^2-\Delta^2)+(q^2_0+q^2+\Delta^2)^2}, \\
 \Delta &=& - 2\Delta g \int\frac{d^3Q}{(2\pi)^3} ~\frac{q^2_0+q^2-m^2+\Delta^2}{m^4 + 2m^2(q^2_0+q^2-\Delta^2)+(q^2_0+q^2+\Delta^2)^2},   
\end{eqnarray}
which upon integrating the frequency and angles becomes
\begin{eqnarray}
    m   &=& \frac{g}{8\pi}\int_0^{\Lambda^2}~dq^2~\left[ \frac{m-\Delta}{\sqrt{q^2+(m-\Delta)^2}} + \frac{m+\Delta}{\sqrt{q^2+(m+\Delta)^2}} \right],\\
 \Delta &=& \frac{g}{8\pi}\int_0^{\Lambda^2}~dq^2~\left[ \frac{m-\Delta}{\sqrt{q^2+(m-\Delta)^2}} - \frac{m+\Delta}{\sqrt{q^2+(m+\Delta)^2}} \right].    
\end{eqnarray}
Summing and subtracting them we get 
\begin{eqnarray}
m+\Delta &=& \frac{g}{4\pi} \int_0^{\Lambda^2}~dq^2\frac{m-\Delta}{\sqrt{q^2+(m-\Delta)^2}},\\
m-\Delta &=& \frac{g}{4\pi} \int^{\Lambda^2}_0~dq^2\frac{m+\Delta}{\sqrt{q^2+(m+\Delta)^2}}.  
\end{eqnarray}
While the limit $\Delta=0,\;\;m\neq0$ is well defined, another limit $m=0,\;\;\;\Delta\neq0$ is not (left hand side is negative, right hand side is positive) which suggests that the $\Delta$-phase cannot exist alone. It remains to check the case of coexistence of both parameters, which arises as the solution of 
\begin{equation}
m\pm\Delta = \gamma(m\mp\Delta)\left[\sqrt{1+(m\mp\Delta)^2} - |m\mp\Delta|\right],
\end{equation}
which appears after integrating the momenta and rescaling all quantities as $m\to \Lambda m,\;\;\Delta\to\Lambda\Delta,\;\;\; \gamma=g\Lambda/2\pi$. The expression in the squared brackets is positive, i.~e. the full equation has any sense only if the sign of $m\pm\Delta$ and $m\mp \Delta$ are the same, which for positive $m$ and $\Delta$ suggests $|m|>|\Delta|$. Below we assume both sides to be positive . 
Rewriting it as 
\begin{equation}
\frac{m\pm\Delta}{\gamma} + (m\mp\Delta)^2 = (m\mp\Delta)\sqrt{1+(m\mp\Delta)^2},
\end{equation}
and squaring both sides we get 
\begin{equation}
\frac{1}{\gamma^2}(m\pm\Delta)^2 = (m \mp \Delta)^2\left[1 - \frac{2}{\gamma}(m\pm\Delta) \right].
\end{equation}
Separating terms which are unique in the sign from those appearing with $\pm$ we find
\begin{equation}
\frac{m^2+\Delta^2}{\gamma^2} - \left(1-\frac{2m}{\gamma}\right)\left(m^2+\Delta^2\right)-\frac{4m\Delta^2}{\gamma^2} =
\mp\frac{2\Delta}{\gamma}\left(\frac{m}{\gamma}+m^2+\Delta^2 +\gamma m\left(1-2\frac{m}{\gamma} \right) \right),
\end{equation}
which can only make sense if both sides vanish individually. Left hand side yields 
\begin{equation}
\Delta^2 = - m^2 \frac{1-\gamma^2+2\gamma m}{1-\gamma^2-2\gamma m},
\end{equation}
while right hand side gives
\begin{equation}
\Delta^2 = -\frac{m}{\gamma}[1+\gamma^2-m\gamma]. 
\end{equation}
Equating both right hand sides yields 
\begin{equation}
|m| = \frac{1-\gamma^4}{4\gamma}, 
\end{equation}
i.~e. it is semi-positive if $\gamma\leqslant1$ and predicts a non-physical phase for small $\gamma$. Moreover, plugging it back yields for $\Delta$
\begin{equation}
\Delta^2 = - \frac{1-\gamma^4}{16\gamma^2} \left[\gamma^4+4\gamma^2+3 \right], 
\end{equation}
which is negative for $\gamma\leqslant1$. But this means that $\Delta$ is imaginary which conflicts with the requirement for the effective Hamiltonian to be hermitian. 
Hence the order parameter $\Delta^{}_2$ must be excluded from the further consideration. 

\section{Structure of the Proca term}
\label{app:Proca}

\no
The Gaussian action is defined by Eq.~(19) in the main part
\begin{equation}
{\cal S}^{}_G[{\cal Q}] =  \vec{\cal Q}^{}_r\cdot{\bf \Pi}^{-1}_{rr^\prime}\vec{\cal Q}^{}_{r^\prime},
\end{equation}
where the real vector field $ \vec{\cal Q}^{}_r = (\vec{\cal Q}^{}_1, \vec{\cal Q}^{}_2)^{\rm T}_r$
\begin{equation}
\vec{\cal Q}^{}_{i} = \left({\cal Q}^{00}_i,{\cal Q}^{01}_i,{\cal Q}^{02}_i,{\cal Q}^{03}_i, 
{\cal Q}^{10}_i,{\cal Q}^{11}_i,{\cal Q}^{12}_i,{\cal Q}^{13}_i,{\cal Q}^{20}_i,{\cal Q}^{21}_i,
{\cal Q}^{22}_i,{\cal Q}^{23}_i,{\cal Q}^{30}_i,{\cal Q}^{31}_i,{\cal Q}^{32}_i,{\cal Q}^{33}_i\right), 
\;\; i=1,2.
\end{equation}
The  matrix ${\cal D}^{-1}_{rr^\prime}$ is different for each phase. Below we give Proca matrices for all phases. 

\begin{enumerate}
\item  $m$-phase:
\begin{eqnarray}
\nn
{\bf \Pi}^{-1}(Q=0) 
&=& M^{}_1\left(e^{}_{1,1} + e^{}_{4,4} + e^{}_{10,10} + e^{}_{11,11} + e^{}_{17,17} + e^{}_{20,20} + e^{}_{26,26} + e^{}_{27,27}\right) \\
\nn
&+& M^{}_2\left(e^{}_{2,2} + e^{}_{9,9} + e^{}_{24,24} + e^{}_{31,31}\right) \\
\nn
&+& M^{}_3\left(e^{}_{3,3} + e^{}_{8,8} + e^{}_{12,12} + e^{}_{15,15} + e^{}_{18,18} + e^{}_{21,21} + e^{}_{25,25} + e^{}_{30,30}\right) \\
\nn
&+& M^{}_4\left(e^{}_{5,5} + e^{}_{14,14} + e^{}_{19,19} + e^{}_{28,28}\right) \\
\nn
&+& M^{}_5\left(e^{}_{6,6} + e^{}_{7,7} + e^{}_{13,13} + e^{}_{16,16} + e^{}_{22,22} + e^{}_{23,23} + e^{}_{29,29} + e^{}_{32,32}\right) \\
\nn
&-& iD^{}_1\left(e^{}_{1,32} + e^{}_{4,29} + e^{}_{6,27} - e^{}_{7,26} + e^{}_{10,23} - e^{}_{11,22} - e^{}_{13,20} - e^{}_{16,17}  \right. \\
\nn
&-&             \left.e^{}_{17,16} - e^{}_{20,13} - e^{}_{22,11} + e^{}_{23,10} - e^{}_{26,7} + e^{}_{27,6} + e^{}_{29,4} + e^{}_{32,1}\right) \\
\nn
&-& D^{}_2\left(e^{}_{2,31} - e^{}_{5,28} - e^{}_{9,24} - e^{}_{14,19} - e^{}_{19,14} - e^{}_{24,9} - e^{}_{28,5} + e^{}_{31,2} \right).
\end{eqnarray}
\no
Explicitly, the matrix elements are 
\begin{equation}
\label{eq:Masses_Mph}
\begin{array}{lll}
\displaystyle M^{}_1 = \frac{2}{\gamma} +  \frac{1}{2\sqrt{1+m^2}},             & \hspace{4mm}\displaystyle   \displaystyle M^{}_2 = \frac{2}{\gamma} + 2m - \frac{1+2m^2}{\sqrt{1+m^2}},  & \hspace{4mm}\displaystyle M^{}_3 = \frac{2}{\gamma},  \\
\displaystyle   M^{}_4 = \frac{2}{\gamma} -  2m + \frac{1+2m^2}{\sqrt{1+m^2}},  & \hspace{4mm}\displaystyle  M^{}_5 = \frac{2}{\gamma} -   \frac{1}{2\sqrt{1+m^2}},                        &           \\
\displaystyle   D^{}_1 = \frac{1}{2\sqrt{1+m^2}}, &\hspace{4mm} \displaystyle  D^{}_2 = -2m + \frac{1+2m^2}{2\sqrt{1+m^2}}. &                 \\
\end{array}
\end{equation}
The difference in comparison to representation in Ref.~\cite{Phonon19} is explained by the fact that some of the matrix elements given there happen to vanish by integration. 
The matrix elements fulfill following equalities:
\begin{eqnarray}
\label{eq:Ward1}
\displaystyle M^{}_1 + M^{}_5 = \frac{4}{\gamma}, &  & \displaystyle  M^{}_1 - D^{}_1 = \frac{2}{\gamma} = M^{}_5 + D^{}_1 \\
\label{eq:Ward2}
\displaystyle M^{}_2 + M^{}_4 = \frac{4}{\gamma}, &  & \displaystyle  M^{}_2 + D^{}_2 = \frac{2}{\gamma} = M^{}_4 - D^{}_2. 
\end{eqnarray}
The eigenvalues of the Gauss matrix for this case are
\begin{equation}
{\rm E}^{m}_{1\cdots 28} = \frac{2}{\gamma},\;\;\;\;  {\rm E}^{m}_{29,30} = \frac{2(3+\gamma^2)}{\gamma+\gamma^3},\;\;\; {\rm E}^{m}_{31,32} = \frac{2(\gamma^2-1)}{\gamma+\gamma^3}.
\end{equation}

\item $\Delta$--phase:
\begin{eqnarray}
\nn
{\bf \Pi}^{-1}(Q=0) 
&=& M^{}_1\left(e^{}_{1,1} + e^{}_{10,10} + e^{}_{17,17} + e^{}_{26,26}\right) + M^{}_2\left(e^{}_{2,2} + e^{}_{9,9}   + e^{}_{24,24} + e^{}_{31,31} \right) \\
\nn
&+& M^{}_3\left(e^{}_{3,3} + e^{}_{12,12} + e^{}_{21,21} + e^{}_{30,30}\right) + M^{}_4\left(e^{}_{4,4} + e^{}_{11,11} + e^{}_{20,20} + e^{}_{27,27} \right) \\
\nn
&+& M^{}_5\left(e^{}_{5,5} + e^{}_{14,14} + e^{}_{19,19} + e^{}_{28,28}\right) + M^{}_6\left(e^{}_{6,6} + e^{}_{13,13} + e^{}_{22,22} + e^{}_{29,29} \right) \\
\nn
&+& M^{}_7\left(e^{}_{7,7} + e^{}_{16,16} + e^{}_{23,23} + e^{}_{32,32}\right) + M^{}_8\left(e^{}_{8,8} + e^{}_{15,15} + e^{}_{18,18} + e^{}_{25,25} \right) \\
\nn
&+& iD^{}_1\left(e^{}_{1,32} - e^{}_{7,26} + e^{}_{10,23} - e^{}_{16,17} - e^{}_{17,16} + e^{}_{23,10} - e^{}_{26,7} + e^{}_{32,1} \right) \\
\nn
&+&  D^{}_2\left(e^{}_{2,31} - e^{}_{9,24} - e^{}_{24,9} + e^{}_{31,2} \right)  + D^{}_3\left(e^{}_{3,30} + e^{}_{12,21} + e^{}_{21,12} + e^{}_{30,3} \right) \\
\nn
&-& iD^{}_4\left(e^{}_{4,29} + e^{}_{6,27} - e^{}_{11,22} - e^{}_{13,20} - e^{}_{20,13} - e^{}_{22,11} + e^{}_{27,6} + e^{}_{29,4} \right) \\
\nn
&+&  D^{}_5\left(e^{}_{5,28} + e^{}_{14,19} + e^{}_{19,14} + e^{}_{28,5} \right).
\end{eqnarray}
\no
The matrix elements are 
\begin{equation}
\label{eq:Masses_Dph}
\begin{array}{lll}
\displaystyle M^{}_1 = \frac{2}{\gamma} +  \frac{1}{2\sqrt{1+\Delta^2}}, 
& \hspace{2mm} \displaystyle M^{}_2 = \frac{2}{\gamma} + 2\Delta - \frac{1+2\Delta^2}{\sqrt{1+\Delta^2}}, 
& \hspace{2mm}   \displaystyle M^{}_3 = \frac{2}{\gamma} -  \Delta + \frac{\Delta^2}{\sqrt{1+\Delta^2}} ,  \\
\displaystyle M^{}_4 = \frac{2}{\gamma} -  \Delta + \frac{1+2\Delta^2}{2\sqrt{1+\Delta^2}}, 
& \hspace{2mm} \displaystyle M^{}_5 = \frac{2}{\gamma} -  \Delta + \sqrt{1 + \Delta^2}, 
& \hspace{2mm}  \displaystyle  M^{}_6 = \frac{2}{\gamma} +  \Delta - \frac{1+2\Delta^2}{2\sqrt{1+\Delta^2}}, \\
\displaystyle M^{}_7 = \frac{2}{\gamma} - \frac{1}{2\sqrt{1+\Delta^2}}, 
& \hspace{2mm} \displaystyle M^{}_8 = \frac{2}{\gamma},             \\
\displaystyle D^{}_1 = -\frac{1}{2\sqrt{1+\Delta^2}}, 
& \hspace{2mm} \displaystyle   D^{}_2 = 2\Delta - \frac{1+2\Delta^2}{2\sqrt{1+\Delta^2}}, 
& \hspace{2mm} \displaystyle D^{}_3 =  \Delta - \frac{\Delta^2}{\sqrt{1+\Delta^2}},\\ 
\displaystyle D^{}_4 = -\Delta + \frac{1+2\Delta^2}{2\sqrt{1+\Delta^2}},  
& \hspace{2mm} \displaystyle   D^{}_5 =  \sqrt{1+\Delta^2} - \Delta,  
&    \\
\end{array}
\end{equation}
We notice following most important equalities which are fulfilled by the matrix elements:
\begin{eqnarray}
\displaystyle M^{}_1 + M^{}_7 = \frac{4}{\gamma}, &  & \displaystyle  M^{}_1 + D^{}_1 = \frac{2}{\gamma} = M^{}_7 - D^{}_1 \\
\displaystyle M^{}_4 + M^{}_6 = \frac{4}{\gamma}, &  & \displaystyle  M^{}_2 - D^{}_2 = \frac{2}{\gamma} = M^{}_4 - D^{}_4. 
\end{eqnarray}

The stability matrix has the following eigenvalues :
\begin{equation}
{\rm E}^{\Delta}_{1\cdots 26} = \frac{2}{\gamma},\;\;\;\; {\rm E}^{\Delta}_{27,28} = \frac{4}{\gamma}, \;\;\;\; {\rm E}^{\Delta}_{29,30} = \frac{4}{\gamma+\gamma^3},\;\;\; {\rm E}^{\Delta}_{31,32} = \frac{2(\gamma^2-1)}{\gamma+\gamma^3}.
\end{equation}
Last two eigenvalues are related to the order parameter of the phase $\Delta$ in Eq.~(17) in the main part. 

\item Coexisting phase $m=\Delta$:
\begin{eqnarray}
\nn
{\bf \Pi}^{-1}(Q=0) 
&=& M^{}_1\left(e^{}_{1,1} + e^{}_{10,10} + e^{}_{17,17} + e^{}_{26,26}\right) + M^{}_2\left(e^{}_{2,2} + e^{}_{9,9}   + e^{}_{24,24} + e^{}_{31,31}\right) \\
\nn
&+& M^{}_3\left(e^{}_{3,3} + e^{}_{12,12} + e^{}_{21,21} + e^{}_{30,30}\right) + M^{}_4\left(e^{}_{4,4} + e^{}_{11,11} + e^{}_{20,20} + e^{}_{27,27}\right) \\
\nn
&+& M^{}_5\left(e^{}_{5,5} + e^{}_{14,14} + e^{}_{19,19} + e^{}_{28,28}\right) + M^{}_6\left(e^{}_{6,6} + e^{}_{13,13} + e^{}_{22,22} + e^{}_{29,29}\right) \\
\nn
&+& M^{}_7\left(e^{}_{7,7} + e^{}_{16,16} + e^{}_{23,23} + e^{}_{32,32}\right) + M^{}_8\left(e^{}_{8,8} + e^{}_{15,15} + e^{}_{18,18} + e^{}_{25,25}\right) \\
\nn
&+& A^{}_1
    \left( e^{}_{1,10} - e^{}_{7,16}  +  e^{}_{10,1}  -  e^{}_{16,7} - ie^{}_{1,23} + ie^{}_{7,17} - ie^{}_{10,32} + ie^{}_{16,26}\right. \\
    \nn
&+& \left.ie^{}_{17,7} - ie^{}_{23,1} + ie^{}_{26,16} - ie^{}_{32,10} + e^{}_{17,26} - e^{}_{23,32} + e^{}_{26,17} - e^{}_{32,23}\right) \\
\nn
&+& A^{}_2\left(e^{}_{2,9} + e^{}_{9,2} - e^{}_{2,24} + e^{}_{9,31} - e^{}_{24,2} + e^{}_{31,9} - e^{}_{24,31} - e^{}_{31,24}\right) \\
\nn
&+& A^{}_3\left(e^{}_{3,30} + e^{}_{12,21} + e^{}_{21,12} + e^{}_{30,3}\right) \\
\nn
&+& iD^{}_1\left(e^{}_{1,32} - e^{}_{7,26} + e^{}_{10,23} - e^{}_{16,17} - e^{}_{17,16} + e^{}_{23,10} - e^{}_{26,7} + e^{}_{32,1}\right) \\
\nn
&+& D^{}_2\left(e^{}_{2,31} - e^{}_{9,24} - e^{}_{24,9} + e^{}_{31,2}\right) \\
\nn
&-& iD^{}_3\left(e^{}_{4,29} + e^{}_{6,27} - e^{}_{11,22} - e^{}_{13,20} - e^{}_{20,13} - e^{}_{22,11} + e^{}_{27,6} + e^{}_{29,4}\right) \\
\nn
&+& D^{}_4\left(e^{}_{5,28} + e^{}_{14,19} + e^{}_{19,14} + e^{}_{28,5}\right).
\end{eqnarray}

Each quantity in the Proca matrix means
\begin{eqnarray}
\nn
A^{}_1 &=& \frac{1-\sqrt{1+4m^2}}{4\sqrt{1+4m^2}}, \\
\nn
A^{}_2 &=& \frac{1}{2}\left[1+4m-2\sqrt{1+4m^2} +\frac{1}{\sqrt{1+4m^2}}\right], \\
\nn
A^{}_3 &=& -\frac{1}{6m^2}\left[1-\sqrt{1+4m^2}  + 2m^2\left(\sqrt{1+4m^2}-2m\right)\right]. 
\end{eqnarray}
Matrix elements $A^{}_{1\cdots3}$ appear only as consequence of coexisting order parameters $m$ and $\Delta$.

\begin{eqnarray}
\nn
M^{}_1 &=& \frac{2}{\gamma} + \frac{1+\sqrt{1+4m^2}}{4\sqrt{1+4m^2}},    \\
\nn
M^{}_2 &=& \frac{2}{\gamma} + \frac{1}{2}\left[4m-1-2\sqrt{1+4m^2} +\frac{1}{\sqrt{1+4m^2}}\right], \\
\nn
M^{}_3 &=& \frac{2}{\gamma} + \frac{1}{6m^2}\left[1-\sqrt{1+4m^2}  + 2m^2\left(\sqrt{1+4m^2}-2m\right)\right],  \\  
\nn
M^{}_4 &=& \frac{2}{\gamma} - \frac{1}{12m^2}\left[1-\sqrt{1+4m^2} - 4m^2\left(\sqrt{1+4m^2}-2m\right)\right], \\
\nn
M^{}_5 &=& \frac{2}{\gamma} -2m + \sqrt{1+4m^2}, \\
\nn
M^{}_6 &=& \frac{2}{\gamma} + \frac{1}{12m^2}\left[1-\sqrt{1+4m^2} - 4m^2\left(\sqrt{1+4m^2}-2m\right)\right], \\
\nn
M^{}_7 &=& \frac{2}{\gamma} -\frac{1+\sqrt{1+4m^2}}{4\sqrt{1+4m^2}},\\
\nn
M^{}_8 &=& \frac{2}{\gamma}. 
\end{eqnarray}
where (and further on) $m$ is defined in Eq.~(17) in the main part.

\begin{eqnarray}
\nn
D^{}_1 &=&  -\frac{1+\sqrt{1+4m^2}}{4\sqrt{1+4m^2}}, \\
\nn
D^{}_2 &=&  \frac{1}{2}\left[4m-1-2\sqrt{1+4m^2} +\frac{1}{\sqrt{1+4m^2}}\right], \\
\nn
D^{}_3 &=& - \frac{1}{12m^2}\left[1-\sqrt{1+4m^2} - 4m^2\left(\sqrt{1+4m^2}-2m\right)\right], \\
\nn
D^{}_4 &=& \sqrt{1+4m^2} - 2m.
\end{eqnarray}
The stability matrix has following  eigenvalues
\begin{eqnarray}
{\rm E}^{}_{1\cdots26} = \frac{2}{\gamma} ,\;\;\;\; {\rm E}^{}_{27,28} = \frac{4}{\gamma},\;\;\;\; {\rm E}^{}_{29,30} = \frac{8}{3\gamma}\frac{1+2\gamma}{(1+\gamma)^2},\;\;\;\;{\rm E}^{}_{31} = \frac{2(\gamma^2-1)}{\gamma+\gamma^3} 
\end{eqnarray}
the latter related to the order parameter defined in Eq.~(17) in the main text. However, the last remaining eigenvalue
\begin{equation}
{\rm E}^{}_{32} = -2 + \frac{2}{\gamma}
\end{equation}
is negative for all $\gamma>1$. 

\end{enumerate}

\section{Evaluation of the current-current correlator}
\label{app:JJ}

Here we evaluate Eq.~(\ref{eq:npKubo}) for the case of usual electronic currents. This correlator is zero in the $\Delta$-phase but 
finite in the $m$-phase, i.e. we have to chose $c=d=1$ and $i=1,\;j=2$
\begin{equation}
\frac{1}{4\gamma^2}\left( {\bf \Lambda}^1_{1,r} {\bf \Lambda}^1_{2,r^\prime}\right) =  
\frac{4}{\gamma^2}\left(
i{\cal Q}^{12}_{1r}{\cal Q}^{21}_{1r^\prime} - i{\cal Q}^{21}_{2r}{\cal Q}^{12}_{2r^\prime} - {\cal Q}^{12}_{1r}{\cal Q}^{12}_{2r^\prime} - {\cal Q}^{21}_{2r}{\cal Q}^{21}_{1r^\prime}
\right). 
\end{equation}
Performing the functional integration we get 
\begin{eqnarray}
\label{eq:HallC}
\frac{1}{4\gamma^2}\langle{\bf \Lambda}^1_{1,r} {\bf \Lambda}^1_{2,r^\prime}\rangle^{}_{\cal G} = \frac{4}{\gamma^2}\int\frac{d^3Q}{(2\pi)^2}~e^{iQ(r-r')}
\frac{1}{2}
\left[{\bf \Pi}^{}_{26,10}(q^{}_0) + {\bf \Pi}^{}_{7,23}(q^{}_0) + i{\bf \Pi}^{}_{26,23}(q^{}_0)  - i{\bf \Pi}^{}_{7,10}(q^{}_0)\right],
\end{eqnarray}
where the factor $1/2$ appears because of the Gauss integral, and the propagator is defined as inverse kernel matrix from Eq.~(\ref{eq:Gauss2})
\begin{equation}
{\bf \Pi}(Q) = 
\left( 
\begin{array}{cc}
 {\bf A} & {\bf B} \\
 {\bf C} & {\bf D}
\end{array}
\right)^{-1}_Q.
\end{equation} 
In $m$-phase we get to linear order in $q^{}_0$ via a direct evaluation 
\begin{eqnarray}
\frac{2}{\gamma^2}\left[{\bf \Pi}^{}_{26,10}(q^{}_0) + {\bf \Pi}^{}_{7,23}(q^{}_0) + i{\bf \Pi}^{}_{26,23}(q^{}_0)  - i{\bf \Pi}^{}_{7,10}(q^{}_0)\right] 
&\approx& -{\rm  sgn}(m)\frac{2}{\gamma^2}\frac{q^{}_0}{4\pi} \frac{(M^{}_1+M^{}_5)^2}{(M^{}_1M^{}_2+\Delta^2)^2}\\
&=& -{\rm  sgn}(m)\frac{q^{}_0}{2\pi},
\end{eqnarray}
where the equalities Eq.~(\ref{eq:Ward1}) were used. Hence we finally obtain for the correlator
\begin{equation}
\frac{1}{4\gamma^2}\langle{\bf \Lambda}^1_{1,r} {\bf \Lambda}^1_{2,r^\prime}\rangle^{}_{\cal G} \approx -\frac{{\rm sgn}(m)}{2\pi}\int\frac{d^3Q}{(2\pi)^2}~e^{iQ(r-r')}~q^{}_0 = 
\frac{{\rm sgn}(m)}{2\pi}~i\partial^{}_\tau\delta(r-r'). 
\end{equation}
Calculating correlators for all index combinations and bringing them into a compact form we get Eq.~(\ref{eq:npKubo}).

\end{document}